\documentclass[showpacs, prb,twocolumn,preprintnumbers , superscriptaddress, aps]{revtex4-2}

\usepackage{color}
\usepackage{amsmath,amssymb}
\usepackage{pifont}
\usepackage{amssymb}  
\usepackage{bbold}
\usepackage{float}
\usepackage{stfloats}
\usepackage[normalem]{ulem}

\usepackage[labelfont=bf]{caption} 
\usepackage{tikz}
\usepackage{makecell}
\usepackage{pifont}   
\usepackage{graphicx} 
\usepackage{dcolumn}  
\usepackage{bm}       
\usepackage{multirow} 
\usepackage{placeins}
\usepackage[colorlinks]{hyperref}
\usepackage{mathtools}
\usepackage{subfigure}

\usepackage[normalem]{ulem}

\newcommand{\vect}[1]{\vect{#1}}
\newcommand{\ket}[1]{\left|{#1}\right\rangle}
\newcommand{\bra}[1]{\left\langle{#1}\right|}

\newcommand{\braket}[2]{\left\langle{#1} | {#2}\right\rangle}

\newcommand{\dket}[1]{\left|{#1}\right\rangle\!\rangle}
\newcommand{\dbra}[1]{\left\langle\!\langle{#1}\right|}
\newcommand{\dbraket}[2]{\left\langle\!\langle{#1} | {#2}\right\rangle\!\rangle}

\usepackage{breqn}
\usepackage{mathrsfs}

\usepackage{multirow}

\begin{document}

\title{Light induced transitions of valley Chern numbers and flat bands in a non-twisted moire graphene-hexagonal boron nitride superlattice}
\author{Saud Alabdulal}
		\affiliation{PSE Division$,$ King Abdullah University of Science and Technology KAUST$,$ Thuwal 23955$,$ Saudi Arabia}
\author{Miftah Hadi Syahputra Anfa}
		\affiliation{Physics Department$,$
		King Fahd University
		of Petroleum $\&$ Minerals$,$
		Dhahran 31261$,$ Saudi Arabia}        
  \author{Hocine Bahlouli}
		\affiliation{Physics Department$,$
		King Fahd University
		of Petroleum $\&$ Minerals$,$
		Dhahran 31261$,$ Saudi Arabia}
      \affiliation{Interdisciplinary Research Center (IRC) for Advanced Materials$,$ KFUPM$,$ Dhahran$,$ Saudi Arabia}
\author{Michael Vogl}
		\affiliation{Physics Department$,$
		King Fahd University
		of Petroleum $\&$ Minerals$,$
		Dhahran 31261$,$ Saudi Arabia}
    \affiliation{Interdisciplinary Research Center (IRC) for Intelligent Secure Systems$,$ KFUPM$,$ Dhahran$,$ Saudi Arabia}
\date{\today}
	\begin{abstract}
 Motivated by the rich topology and interesting quasi-band structure of twisted moire materials subjected to light, we study a non-twisted moire material under the influence of light. Our work is in part motivated by a desire to find an easier-to-synthesize platform that can help experimentally elucidate the interesting physics of moiré materials coupled to light. Similar to twisted moire materials, we uncover rich topology and interesting band flattening effects, which we summarize in relevant plots such as a topological phase diagram. Our work demonstrates that much of the interesting phenomenology of twisted moire materials under the influence of electromagnetic waves seems to be generically present even in more experimentally accessible untwisted moire platforms, which remain highly tunable by light.
 \end{abstract}

\maketitle

\section{Introduction}

Recent advances in high-frequency lasers have paved the way for promising developments in condensed matter physics. This type of laser can be used for probing dynamics of, for example, electrons inside an atom \cite{Hentschel2001,Hu2006}, ferroelectrics \cite{Rana2009,Sheu2014,Zhang2021,Chen2016}, exciton in carbon nanotubes \cite{Stich2013,Bai2018,Birkmeier2022}, and solid-state materials \cite{Jager2017,Bionta2021}. It also provides unprecedented control over solid-state systems through time-domain manipulation, such as controlling standing spin wave dynamics \cite{Deb2022}, magnetism \cite{Graves2013,Stanciu2007}, and electronic phases \cite{Rini2007} in several solids. On the theoretical side, progress in this topic has been motivated by the realization that a system, when driven periodically, remains in a so-called prethermal regime for an exponentially long time rather than being completely disordered instantaneously, even if the system under consideration is interacting \cite{Weidinger2017}. This observation enables us to study such a system employing methods similar to an equilibrium one, via an effective time-independent Hamiltonian \cite{Giovannini2019,Oka2019,Sentef2015,Ito2023,Sentef2020,Kennes2019,Eckhardt2022,Topp2021,VinasBostroem2020,Kibis2020,Kong2022,Rudner2020,Castro2022,Fleury2016,Zhan2023,Shin2018,Cao2024,Grushin2014,Ashida2020,Li2020,Topp2019, Vogl_2020,Vogl2021} 

Another reason why applying high-frequency lasers to solid-state systems is interesting is that it has been predicted to lead to exotic phenomena. Examples include superconducting states, such as those reported in \cite{Mitrano2016, Fausti2011, Suda2015, Fava2024}, where the transition appears to be induced by a pulse of light.  It has also been reported that light can induce ferroelectricity in SrTiO$_3$ \cite{Nova2019, Li2019} and an effective magnetic field in antiferromagnetic ErFeO$_3$ \cite{Nova2016}.

Additional interesting results have also been obtained in the context of 2D materials. Among the most celebrated are the so-called Floquet topological insulating states. This effect has been predicted theoretically in monolayer graphene when driven with circularly polarized light \cite{Ashida2020} and subsequently observed experimentally in \cite{McIver2019}.

Even more interesting effects appear once one couples a laser to multilayer materials with Moir\'{e} patterns. One way to construct such a pattern is via a relative twist between layers. In this context, studies have suggested that light can be used to induce flat energy bands in a twisted bilayer graphene \cite{Li2020} and tune the topological properties of its electronic band structure \cite{Topp2019, Vogl_2020}. Apart from twisted bilayer graphene, the interaction between light and twisted transition metal dichalcogenides (TMDs) has also been studied \cite{Vogl2021}, leading to topological transitions in the energy bands. We note that similar effects also occur in the case of applied external pressure\cite{5k9m-mfbz}, although this work will focus on the effects of external laser fields.

Moir\'{e} patterns, however, are not restricted to twisted materials; instead, they also appear often when there are slight mismatches between lattice structures. For instance, they may also appear by stacking two or more layers that have a slight difference in lattice constants. A well-known example of this type is a hetero-bilayer of graphene and hexagonal boron nitride (G-hBN) \cite{Moon_2014}. Unlike their twisted counterpart, however, non-twisted Moir\'{e} materials under the influence of light have not been exhaustively studied in literature. Therefore, this will be the focus of our current work.

The remainder of the paper is structured as follows. In Section \ref{sec:equilibrium_case}, we will review the case of the G-hBN bilayer in equilibrium, i.e., without the influence of light. We will present the model developed in the literature and illustrate some well-known results to ensure our work is self-contained. Then in Section \ref{sec:nonequilibrium_model}, we prepare the model for the case when the G-hBN bilayer is under the influence of light. We then discuss results from our calculation in Section \ref{sec:nonequilibrium_result}. In Section \ref{sec:concluscion}, we conclude our work with a summary and discussion of future directions.

\section{Brief review of the equilibrium model and its results}
\label{sec:equilibrium_case}
\begin{figure}[H]
    \centering
    \includegraphics[width=0.7\linewidth]{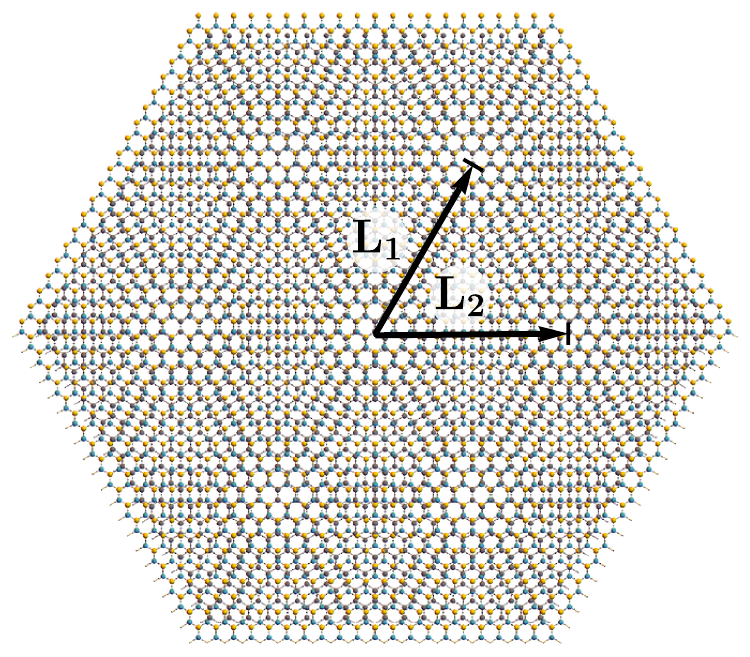}
    \caption{A cartoon that demonstrates the moire pattern of Graphene (gray)/hBN (yellow and blue) and a moire superlattice vector}
    \label{fig:G/hBN Crystal Structure Demonstratoin}
\end{figure}

\begin{figure}[H]
    \centering
    \includegraphics[width=\linewidth]{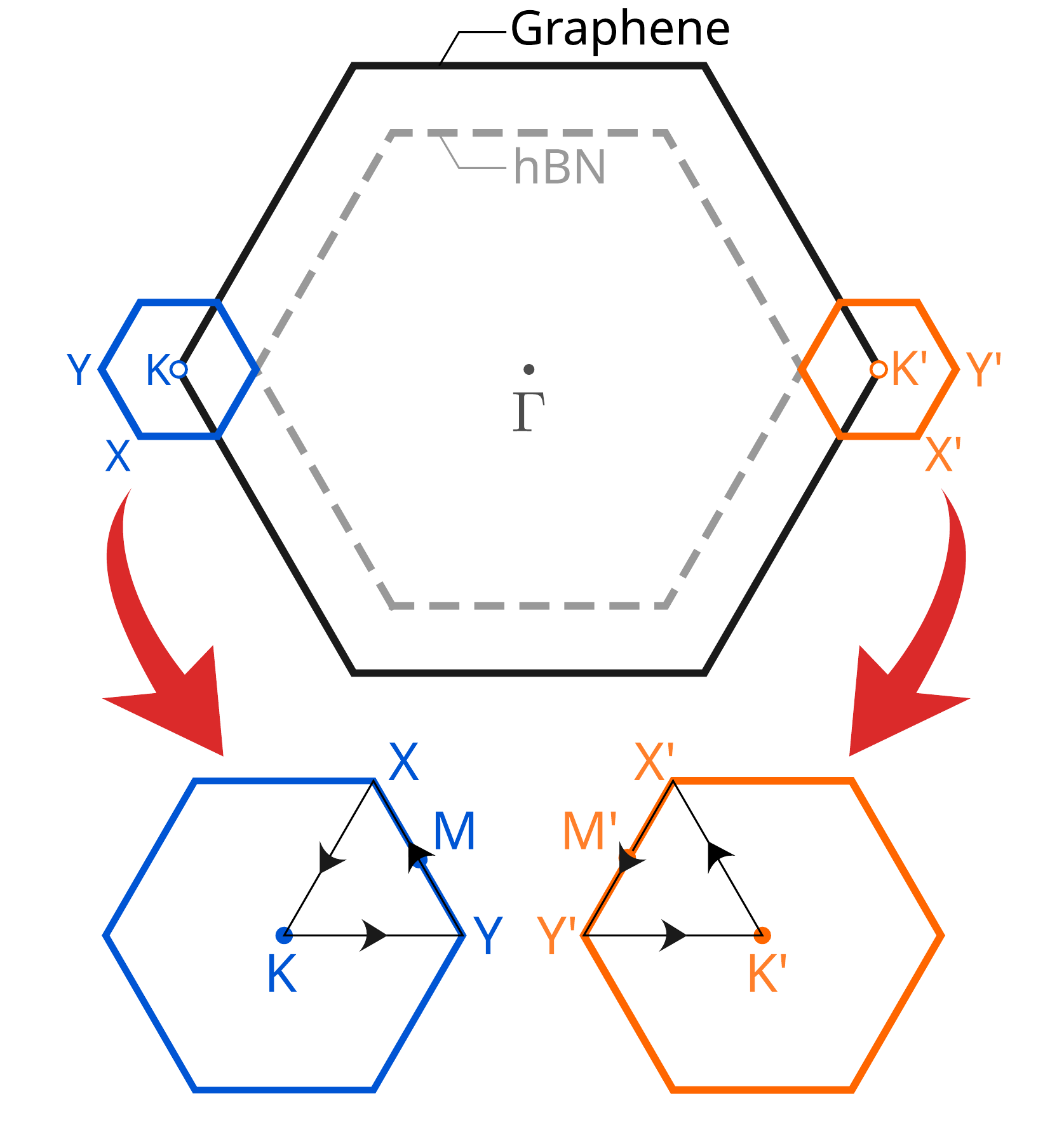}
    \caption{An illustration of the moir\'{e} Brillouin zone (BZ) of both $K$ and $K'$ valley with the path used for the band structure calculation. Note that the ratio between graphene and hBN Brillouin zones has been greatly altered for illustrative purpose.}
    \label{fig:BZ}
\end{figure}

Our study focuses on a bilayer moir \'{e} material consisting of hexagonal Boron Nitride (hBN) and Graphene. A mismatch between lattice constants of this hetero-bilayer structure leads to the emergence of a moir\'{e} superlattice as depicted in Fig. \ref{fig:G/hBN Crystal Structure Demonstratoin}a. Like any periodic structure, it has an associated reciprocal lattice with the corresponding Brillouin zone depicted in Fig. \ref{fig:BZ}.

Both Graphene and hBN have hexagonal symmetry and a two-atom basis. We refer to the sublattices associated with each basis atom as A and B, respectively. In the case of Graphene, both basis atoms are identical (carbon), implying a sublattice symmetry, and in the case of hBN, the two atoms differ (one is Boron and the other Nitrogen) also has a two-atom base hexagonal structure with Boron and Nitrogen atoms. We will consider the case where both lattices are oriented the same way, such that lattice vectors for both Graphene $\mathbf{a}^{\mathrm{G}}_i$ and hBN $\mathbf{a}^{\mathrm{hBN}}_i$ can be defined simultaneously as
\begin{equation}
    \mathbf{a}^{(\gamma)}_1 = a_\gamma (1, 0), \quad \mathbf{a}^{(\gamma)}_2 = a_\gamma(1/2, \sqrt{3}/2)
\end{equation}
where $\gamma = ({\mathrm{G}}, {\mathrm{hBN}})$ stands for graphene and hBN, respectively. We stress that lattice constants differ for the two layers and are given as $a_{\mathrm{G}} \approx 0.2460$ nm \cite{Koshino2013} and $a_{\mathrm{hBN}} \approx 0.2504$ nm \cite{Liu2003}.  It is important to compute reciprocal lattice vectors of Graphene and hBN to be able to find the moir\'{e} reciprocal superlattice vector. Using $\mathbf{a}_i^{(\gamma)} \cdot \mathbf{b}_j^{(\gamma)} = 2\pi\delta_{ij}$, the reciprocal lattice vectors of Graphene and hBN are found to be $\mathbf{b}_1^{(\gamma)} = (1, -1/{\sqrt{3}})2 \pi/{a_{\gamma}}$ and $\mathbf{b}_2^{(\gamma)} = (0, {2}/{\sqrt{3}})2 \pi/{a_{\gamma}}$. To find moir\'{e} superlattice vectors, we recognize that they can be defined as the shortest vectors connecting points where the graphene and hBN lattices are aligned. The number of times Graphene's lattice vector can be repeated between such points is $a_{\mathrm{hBN}}/(a_{\mathrm{G}}-a_{\mathrm{hBN}})$. Since the superlattice orientation in our case is the same as both underlying lattices, the superlattice vectors are then given as
\begin{align}
    \mathbf{L}_i = \frac{a_{\mathrm{hBN}}}{a_{\mathrm{G}}-a_{\mathrm{hBN}}}\mathbf{a}_i
    \label{eq:moire_vectors}
\end{align}

Moir\'{e} reciprocal lattice vectors $\mathbf{G_j}$ can then be obtained by using the relation $\mathbf{L_i}\cdot\mathbf{G_j}=2\pi \delta_{ij}$  and given below
\begin{align}
    \mathbf{G}_i = \frac{a_{\mathrm{G}}-a_{\mathrm{hBN}}}{a_{\mathrm{hBN}}}\mathbf{b}_i.
    \label{eq:moire_reciprocal_vectors}
\end{align}

The Hamiltonian of the system, according to \cite{Moon_2014} and verified by a tight-binding description, is given below
\begin{align}
&H_{\mathrm{G}-\mathrm{hBN}}=\left(\begin{array}{cc}
H_{\mathrm{G}} & U^{\dagger} \\
U & H_{\mathrm{hBN}}
\end{array}\right).
\end{align}
Here,  $H_{\mathrm{G}}(\mathbf{k})$ is the approximate Hamiltonian of a monolayer graphene that is valid near the $\mathbf{K}$ and $\mathbf{K'}$ points and given as 
\begin{equation}
\label{eq:g_hamiltonian}
H_{\mathrm{G}}(\mathbf{k})\approx-\hbar v \mathbf{k}\cdot \boldsymbol{\sigma}_{\xi}, \quad \boldsymbol{\sigma}_{\xi}=\left(\xi \sigma_x, \sigma_y\right),
\end{equation}
where $\xi = +1 \ (-1)$ indicates the $K \ (K')$ valley and  $v=0.8\times 10^6 \mathrm{ m/s}$ is the so-called Fermi velocity of graphene. The second part of the Hamiltonian, $H_{\mathrm{hBN}}$, is an approximate hBN Hamiltonian
\begin{align}
    H_{\mathrm{hBN}} \approx\left(\begin{array}{cc}
V_{\mathrm{N}} & 0 \\
0 & V_{\mathrm{B}}
\end{array}\right)
\end{align}
 where $V_N = -1.40$ eV and $V_B = 3.34$ eV are the potential due to the Nitrogen and Boron, respectively \cite{Slawinska2010}. Lastly, $U$ represents the interlayer hopping term between the graphene and hBN layer and is given as below
\begin{align}
&\begin{aligned}
U= & \left(\begin{array}{ll}
U_{A_{hBN} A_G} & U_{A_{hBN} B_G} \\
U_{B_{hBN} A_G} & U_{B_{hBN} B_G}
\end{array}\right)=u_0\left[\left(\begin{array}{ll}
1 & 1 \\
1 & 1
\end{array}\right)+\right. \\
& \left.\left(\begin{array}{cc}
1 & \zeta_3^{-\xi} \\
\zeta_3^{\xi} & 1
\end{array}\right) e^{i \xi \mathbf{G}_1 \cdot \mathbf{r}}+\left(\begin{array}{cc}
1 & \zeta_3^{\xi} \\
\zeta_3^{-\xi} & 1
\end{array}\right) e^{i \xi\left(\mathbf{G}_1+\mathbf{G}_2\right) \cdot \mathbf{r}}\right],
\end{aligned}
\end{align}
where we used $ \zeta_3=\exp(2\pi i/3)$ as a short-hand notation and  $u_0 \approx 0.152$ eV according to Ref. \cite{Moon_2014}. In detail $U_{ij}$ corresponds to hoppings between specific graphene sites $A_{\mathrm{G}}$ and $B_{\mathrm{G}}$ and hBN sites $A_{\mathrm{hBN}}$  (Nitrogen), $B_{\mathrm{hBN}}$ (Boron).

Since the hBN Hamiltonian $H_{\mathrm{hBN}}$ has a band gap at small energies, physics here is dominated by the Graphene Hamiltonian $H_{\mathrm{G}}$. Therefore, the Hamiltonian of the system using degenerate perturbation theory can be approximated as
\begin{equation}
\label{eq:ghbn_hamiltonian}
H_{\mathrm{G}-\mathrm{hBN}} =  H_{\mathrm{G}}+V_{\mathrm{hBN}},
\end{equation}
where
\begin{align}
\label{eq:hbn_superpotential}
\begin{aligned}
& V_{\mathrm{hBN}}=V_0\left(\begin{array}{ll}
1 & 0 \\
0 & 1
\end{array}\right) \\
&+\left\{V _ { 1 } e ^ { i \xi \psi } \left[\left(\begin{array}{cc}
1 & \zeta_3^{-\xi} \\
1 & \zeta_3^{-\xi}
\end{array}\right) e^{i \xi \mathbf{G}_1 \cdot \mathbf{r}}+\left(\begin{array}{cc}
1 & \zeta_3^{\xi} \\
\zeta_3^{\xi} & \zeta_3^{-\xi}
\end{array}\right) e^{i \xi \mathbf{G}_2 \cdot \mathbf{r}}\right.\right. \\
&\left.\left.+\left(\begin{array}{cc}
1 & 1 \\
\zeta_3^{-\xi} & \zeta_3^{-\xi}
\end{array}\right) e^{-i \xi\left(\mathbf{G}_1+\mathbf{G}_2\right) \cdot \mathbf{r}}\right]+ \text { h.c. }\right\}.
\end{aligned}
\end{align}
is a superpotential due to the hBN layer that electrons in the graphene layer experience.The parameters $V_0$ and $V_1$ are defined as \cite{Moon_2014}
\begin{align}
    V_0 &= -3u_0^2 \left( \frac{1}{V_N} + \frac{1}{V_B} \right), \\
    V_1 &= -u_0^2e^{-i\psi} \left( \frac{1}{V_N} + \zeta_3 \frac{1}{V_B} \right)
\end{align}
In accordance with Ref. \cite{Moon_2014}, we set
$V_0 \approx 0.0289$ eV, $V_1 \approx 0.0210$ eV, and $\psi \approx -0.29$ rad.

As can be seen from the expression of Eqn. \eqref{eq:g_hamiltonian} and \eqref{eq:hbn_superpotential}, the effective  Hamiltonian \eqref{eq:ghbn_hamiltonian} can be used to study the $K$ and $K'$ valleys separately. Using Eqn. \eqref{eq:moire_reciprocal_vectors}, one can construct the moir\'{e} Brillouin zone (BZ) of this system, centered at the $K$ and $K'$ valleys. The illustration of the moir\'{e} BZ and its high-symmetry points are depicted in Figure \ref{fig:BZ}. We note that the location of $K$ ($K'$) valley is given by $\mathbf{K} = -\xi(2\mathbf{b}_1^{(\mathrm{G})} + \mathbf{b}_2^{(\mathrm{G})} )/ 3 $.

\begin{figure}[H]
    \centering
    \includegraphics[width=1.0\linewidth]{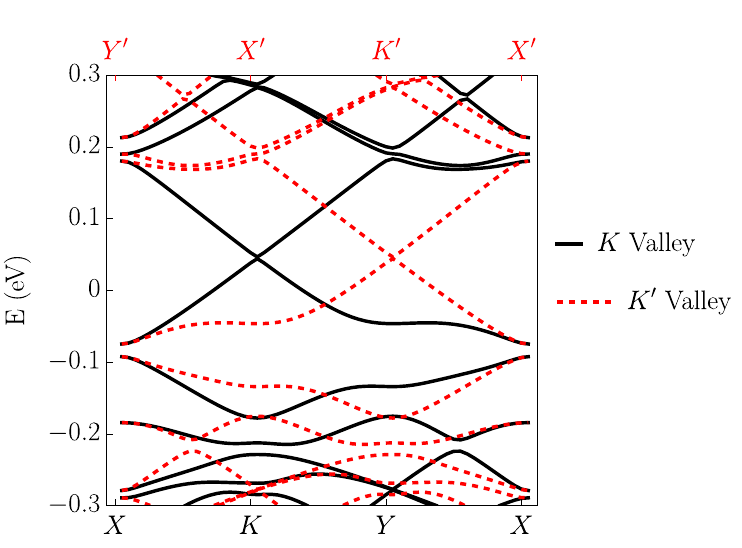}
    \caption{The band structure for hBN/G Moire system for $K,K'$ approximations in black and dashed red respectively.}
    \label{fig: The band Structure of G/hBN}
\end{figure}

In order to calculate the eigenvalues of the Hamiltonian, we need to use Bloch's theorem \cite{bloch1929} to write the Hamiltonian in a plane wave basis. Bloch's theorem allows us to express the wave function of a periodic system as 
\begin{align}
\label{Bloch State}
    \ket{\psi(\mathbf{x})} = e^{i \mathbf{k}\cdot\mathbf{x}} \ket{u(\mathbf{x})}\\
    \ket{u(\mathbf{x}+\mathbf{R})} = \ket{u(\mathbf{x})}.
\end{align}
If we substitute Eq.\eqref{Bloch State} into the Schrödinger equation, we get the Schrödinger equation of Bloch electrons
\begin{align}
    E \ket{u(\mathbf{x})}=H(\mathbf{p}+\mathbf{k}, \mathbf{x}) \ket{u(\mathbf{x})}.
\end{align}
This equation can now be employed to compute the band structure by expanding the Hamiltonian in a plane wave basis $\ket{\alpha}=\exp{(-i\mathbf{G}_\alpha\cdot\mathbf{x})}/\sqrt{\Omega}$ as
\begin{align*}
    \bra{\beta}H_{\text{G-hBN}}(k)\ket{\alpha} = \frac{1}{\Omega} \int e^{i \mathbf{G}_{\beta} \mathbf{x}} H(\mathbf{k}, \mathbf{p}, \mathbf{x}) e^{-i \mathbf{G}_{\alpha} \mathbf{x}} \ \mathrm{d}\mathbf{x},
\end{align*}
where ${\Omega}$ is the area of the real-space unit cell. This form of the Hamiltonian is useful because it allows us to compute the band structure of the Hamiltonian numerically. A computed band structure is shown in Fig. \ref{fig: The band Structure of G/hBN}.

We find that our result agrees with the results in Ref. \cite{Moon_2014}. We observe that the bands near the $\mathbf K$ and $\mathbf K^\prime$ valleys share similar features, as they both exhibit a gapped region and Dirac-cone-like structures.

\section{Non-equilibrium model}
\label{sec:nonequilibrium_model}
\begin{figure}[H]
    \centering
    \includegraphics[width=0.7\linewidth]{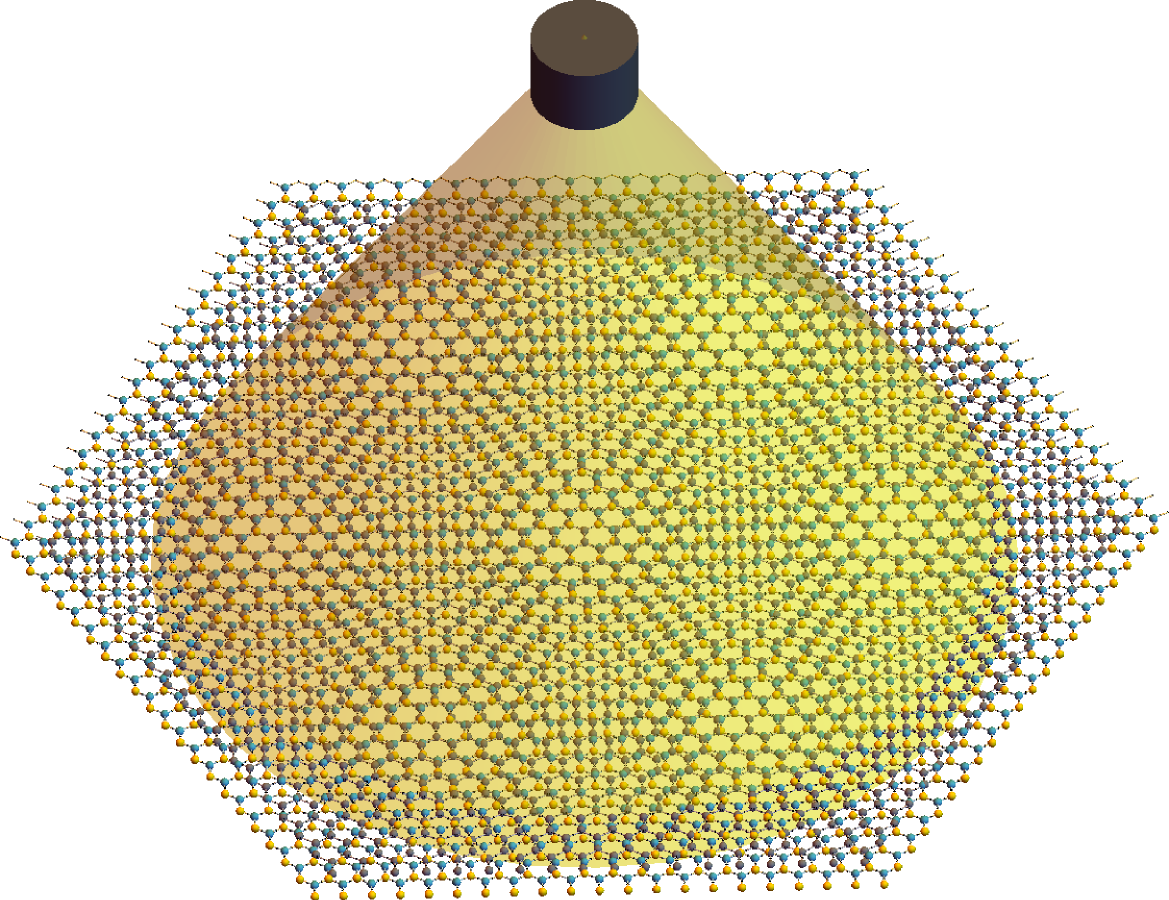}
    \caption{A cartoon demonstration of irradiating a circularly polarized light on the Graphene/hBN system}
    \label{fig:Light on G/hBN}
\end{figure}

Now, we want to observe the effect of circularly polarized light on the Graphene and hBN hetero-bilayer (see Fig. \ref{fig:Light on G/hBN}). We start by considering the general time-dependent Schrodinger equation
\begin{align}
\label{TDSE 1}
    i \partial_t \ket{\psi} = H(t) \ket{\psi}.
\end{align}

for a time-periodic Hamiltonian $H(t)=H(t+T)$, where we may apply Floquet theory (analogous to Bloch theorem \cite{bloch1929} but in time), which allows us to write any solutions in the following form
\begin{align}
\label{Floquet theory equation 1}
&\ket{\psi_n(t)} = e^{-i \varepsilon_n t} \ket{u_n(t)}, \\
\label{Periodic function Fl}
&\ket{u_n(t+T)} = \ket{u_n(t)}.
\end{align}
In the expression above, $\varepsilon_n$ is called quasienergy and is a solution for the eigenvalue problem $Q(t) \ket{u_n(t)} = \varepsilon_n \ket{u_n (t)}$ where $Q(t)=H(t)-i \partial_t$ is the so-called quasienergy operator. This operator acts in a larger space called Sambe space $\mathcal{F}$ which is a direct product of the original Hilbert space and a space of time-periodic functions. For any kets $\dket{u_n}, \dket{u_m} \in \mathcal{F}$, the inner product between them is defined as
\begin{equation}
    \dbraket{u_n}{u_m} = \frac{1}{T} \int_0^T \braket{u_n(t)}{u_m(t)} \, \mathrm{d}t,
    \label{eq:inner_prod}
\end{equation}
where $\braket{\, \cdot \,}{\, \cdot \,}$ is the inner product of the original Hilbert space. This definition is suitable for this space since it takes into account the temporal variables in addition to the spatial variables. 

We choose as our basis the so-called Floquet-Bloch states $\dket{\alpha, m} = \ket{\alpha} \exp(i m\omega_0 t)$, where we use the same definition $\ket{\alpha}=\exp{(-i\mathbf{G}_\alpha\cdot\mathbf{x})}/\sqrt{\Omega}$ as is in Section \ref{sec:equilibrium_case}. Thus, the matrix elements in our chosen basis are given as
\begin{equation}
\begin{aligned}
    Q_{\beta \alpha,nm} &= \dbra{\beta, n}H(t)-i \partial_t \dket{\alpha, m}\\
    &=\dbra{\beta, n}H(t) \dket{\alpha, m}  + m \omega_0 \delta_{\mathbf{G}_{\beta}, \mathbf{G}_{\alpha}} \delta_{n, m}.
\end{aligned}
\label{eq: Floquet Matrix Elements}
\end{equation}
Similar to the case of plane wave expansion of a periodic Hamiltonian using Bloch's theorem, the expression in Eq. \eqref{eq: Floquet Matrix Elements} is useful since it facilitates a numerical treatment.

To account for the effect of circularly polarized light, we make the following replacement $\mathbf{k} \rightarrow \mathbf{k}(t)=\mathbf{k}+\mathbf{A}(t)$ in Eq. \eqref{eq:g_hamiltonian} where
\begin{equation}
\mathbf{A}(t)=A\left( \sin(\omega_0 t)\hat{x}+ \cos(\omega_0 t)\hat{y}\right).
\end{equation}

Since circularly polarized light consists of transverse elements, the effects on interlayer hoppings are considered negligible for this study, and we therefore treat the replacement above as the only effect due to light.

Matrix elements for Eq. \eqref{eq:ghbn_hamiltonian}, i.e. the matrix elements $\dbra{\beta, n} H (t)\dket{\alpha,m}$ can then be found if we decompose $H(t)=H_{G} (\mathbf{k}, t)+V_{hBN}$ to obtain
\begin{widetext}
\begin{equation}
    \dbra{\beta, n} H_{G} (\mathbf{k}, t) \dket{\alpha,m} = -\hbar v \bigg[ (\mathbf{k} + \mathbf{G}_\alpha) \cdot \boldsymbol{\sigma}_{\xi} \delta_{n,m} 
    -  \frac{A}{2} \bigg( -i\xi \sigma_x (\delta_{n,m-1} - \delta_{n,m+1}) 
    +  \sigma_y(\delta_{n,m-1} + \delta_{n,m+1}) \bigg) \bigg] \delta_{\mathbf{G}_\beta, \mathbf{G}_\alpha}     
    \label{eq:Hg_Floquet}
\end{equation}
and
\begin{equation}
    \begin{aligned}
    \dbra{\beta,n}V_{\mathrm{hBN}} \dket{\alpha,m} = \Bigg\{ V_0\left(\begin{array}{ll}
    1 & 0 \\
    0 & 1
    \end{array}\right) \delta_{\mathbf{G}_\alpha, \mathbf{G}_\beta} 
    + \Bigg[ V _ { 1 } e^{ i \xi \psi } \bigg[&\left( \begin{array}{cc}
    1 & \zeta_3^{-\xi} \\
    1 & \zeta_3^{-\xi}
    \end{array}\right) \delta_{\xi\mathbf{G}_1,\mathbf{G}_\alpha - \mathbf{G}_\beta} 
    +\left(\begin{array}{cc}
    1 & \zeta_3^{\xi} \\
    \zeta_3^{\xi} & \zeta_3^{-\xi}
    \end{array}\right) \delta_{\xi\mathbf{G}_2,\mathbf{G}_\alpha - \mathbf{G}_\beta}  \\
    &+ \left(\begin{array}{cc}
    1 & 1 \\
    \zeta_3^{-\xi} & \zeta_3^{-\xi}
    \end{array}\right) \delta_{\xi(\mathbf{G}_1 + \mathbf{G}_2),\mathbf{G}_\alpha - \mathbf{G}_\beta} \bigg]+ \text { h.c. } \Bigg] \Bigg\} \delta_{n,m}.
    \end{aligned}
    \label{eq:VhBN_Floquet}
\end{equation}
\end{widetext}

From here, energy bands can be computed straightforwardly.

However, another primary objective of this study is also to calculate the valley Chern number of the graphene/hexagonal boron nitride (hBN) hetero-bilayer when subjected to the periodic drive and thereby explore its topological properties. Here, we focus on the high-frequency regime because, unlike for low-frequency drives, the Chern number captures the important physics\cite{rudner2013} (eliminating the need for a more complicated winding number analysis). Moreover, at high frequencies, Floquet copies are sufficiently separated, preventing band crossings between different Floquet copies that, in a multi-band analysis, can lead to complications—a chaotic mix of bands. 

The Chern number of the nth band can be expressed in terms of Berry curvature $\mathbf{B}_n(\mathbf{k})$ as follows \cite{Wawer_2021}
\begin{align}
    \mathcal{C}_n=\frac{1}{2 \pi} \oint_{BZ} d \mathbf{k} \cdot \mathbf{B}_n(\mathbf{k}).
    \label{eq:Chern number using BC}
\end{align}
where 
\begin{align}
    &\mathbf{B}_n(\mathbf{k}) =\Delta_{\mathbf{k}} \times \mathcal{A}_n (\mathbf{k}), \\
    &\mathcal{A}_n (\mathbf{k}) = i\bra{\psi_n(\mathbf{k})} \Delta_{\mathbf{k}} \ket{\psi_n(\mathbf{k})}
\end{align}
and $\ket{\psi_n(\mathbf{k})}$ is the normalized eigenstate of the $n$th band.

Calculating the Chern number using obvious discretization approaches of Eq. \eqref{eq:Chern number using BC} can be computationally expensive. To address this, we adopt an efficient method introduced by Fukui et al. \cite{Fukui_2005}, which discretizes the Brillouin zone and utilizes link variables to compute the Chern number more effectively.\\

We begin by discretizing the Brillouin zone into an $N_1\times N_2$ grid of momentum points $k_{\ell}$, where $\ell=\left(\ell_1, \ell_2\right)$ with $\ell_1=$ $1, \ldots, N_1$ and $\ell_2=1, \ldots, N_2$. Discretized points $k_\ell$ are then defined via step sizes $\Delta k_\mu$ for each direction $\mu$ as

\begin{align}
  k_{\ell}=\left(\ell_1 \Delta k_1, \ell_2 \Delta k_2\right); \quad \Delta k_\mu=\frac{X_\mu}{N_\mu}  ,
\end{align}
where $X_\mu$ is the length of the Brillouin zone in the $\mu$-direction, and $N_\mu$ is the number of discretization points along that axis. 

We may now introduce so-called link variables $U_\mu\left(k_{\ell}\right)$ as
\begin{align}
    U_\mu\left(k_{\ell}\right) =\frac{\left\langle n\left(k_{\ell}\right) \mid n\left(k_{\ell}+\hat{\mu}\right)\right\rangle}{\left|\left\langle n\left(k_{\ell}\right) \mid n\left(k_{\ell}+\hat{\mu}\right)\right\rangle\right|},
    \label{link variable}
\end{align}
where $\left|n\left(k_{\ell}\right)\right\rangle$ is the eigenstate of the $nth$ band at the point $k_\ell$. In our expression above, we used vectors $\hat{\mu}=\Delta k_\mu(\delta_{1\mu},\delta_{2\mu})$ that point from one discretized site to another.  

Next, following \cite{Fukui_2005}, we define the lattice field strength $F_{12}\left(k_{\ell}\right)$ on each point of the discretized Brillouin zone:
\begin{align}
    F_{12}\left(k_{\ell}\right)=\ln \left[U_1\left(k_{\ell}\right) U_2\left(k_{\ell}+\hat{1}\right) U_1^{-1}\left(k_{\ell}+\hat{2}\right) U_2^{-1}\left(k_{\ell}\right)\right]
\end{align}

Finally, according to \cite{Fukui_2005}, the Chern number $C_n$ for the $n$-th band is calculated by summing the field strength over all points in the Brillouin zone
\begin{align}
     C_n = \frac{1}{2 \pi i} \sum_{\ell} F_{12}\left(k_{\ell}\right).
\end{align}

With all pieces in place, we are now in the position to present results in the next section.

\section{Non-equilibrium results}
\label{sec:nonequilibrium_result}
\begin{figure}[H]
    \centering
    \includegraphics[width=\linewidth]{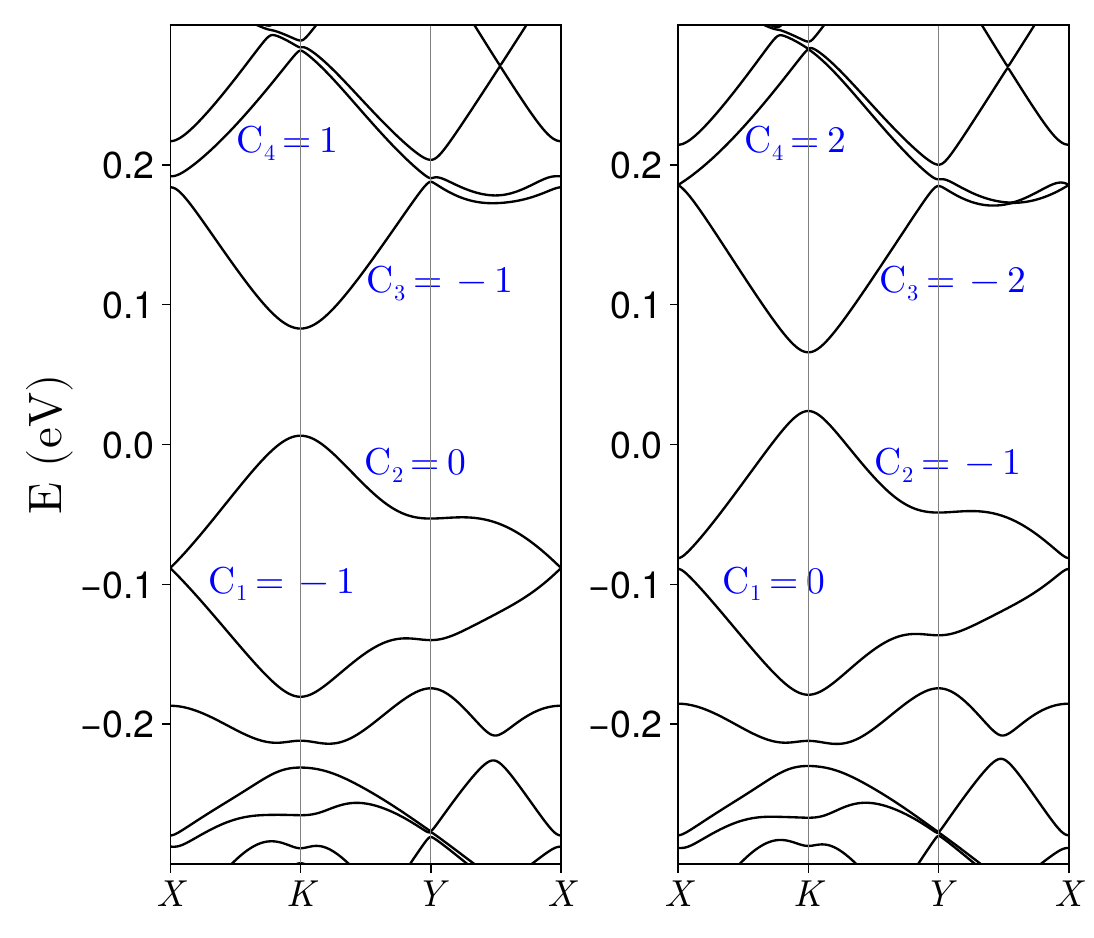}
    \caption{The $K$-valley band structure of G-hBN under the influence of circularly polarized light for (left) $\omega_0=4.5$ and (right) 8.5 eV at fixed $Aa_0 = 0.2$. The Chern numbers for the low-energy hole (electron) bands are also shown. We stress that, although due to resolution, it may appear that bands 3 and 4 overlap, they do not, which makes the computed Chern numbers well-defined.}
    \label{fig:band_Aj0.2}
\end{figure}

We begin with an investigation of the quasi-energy band structure of the G-hBN hetero bi-layer, which can be obtained by diagonalizing the Floquet Hamiltonian Eq. \eqref{eq: Floquet Matrix Elements}. For our analysis, we truncated to three Floquet copies, i.e., indices run as $m,n=-1,0,1$. This simplification restricts us to a relatively high frequency regime. Fig. \ref{fig:band_Aj0.2} shows the quasi-energy band structure of the G-hBN (for only the $K$-valley) system at a fixed driving strength $A=0.2/a_0$ for driving frequency $\omega_0=4.5$ and 8.5 eV. We restricted our analysis to the $\mathbf{K}$ valley to avoid overloading the figure; the $\mathbf{K}^\prime$ valley experiences similar physics.
Chern numbers are also labeled for four bands (labeled 1 through 4, as seen in the figure), and we observe that they can be modified by adjusting the driving frequencies. We note that although visually it is suggested that bands 3 and 4 are crossing for the case of $\omega_0=8.5$ eV due to resolution, there is actually a gap of $\approx 0.34$ meV between these two bands.Moreover, as shown in the plot, the application of circularly polarized light modifies various band gaps, such as the one between bands 2 and 3. 

\begin{figure}[H]
    \centering
    \includegraphics[width=\linewidth]{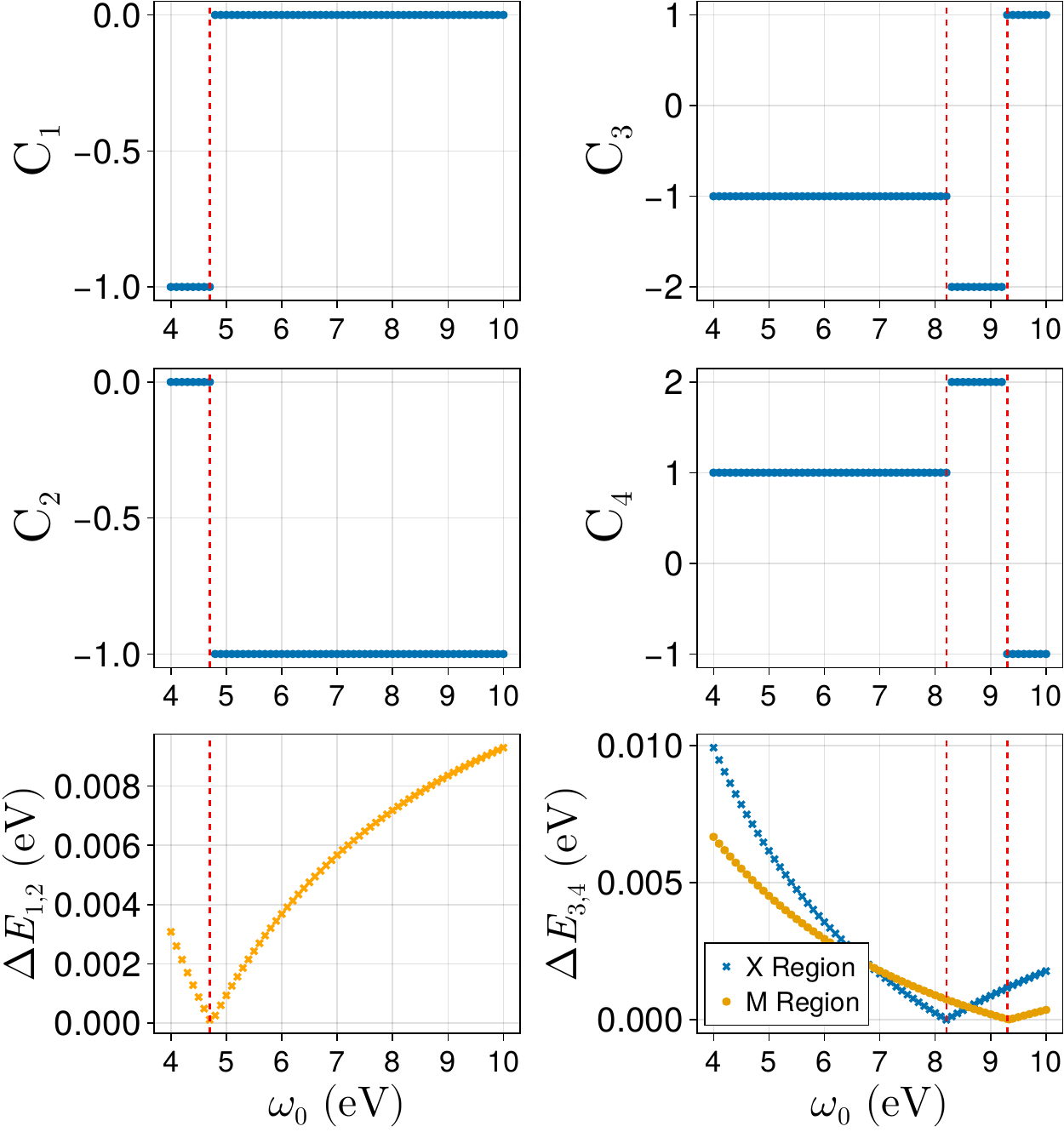}
    \caption{Chern numbers of four low energy bands for a fixed driving strength $Aa_0 = 0.2$ and as a function of driving frequency $\omega_0$. The two lower plots show the energy gap between the two pairs of bands.}
    \label{fig:chern_gap}
\end{figure}

 Changes in frequency $\omega_0$ - even at fixed driving strength $Aa_0$ - can lead to band gap closings, which permit a change in Chern number. Such a change is interpreted as a topological phase transition. Fig. \ref{fig:chern_gap} shows the Chern numbers of the four bands as marked in Fig. \ref{fig:band_Aj0.2} (we consider the $K$ valley only - results for the $K^\prime$ valley are similar). Here, our computations were done as a function of driving frequency $\omega_0$ at fixed driving strength $A = 0.2/a_0$. The two lower plots show the energy gaps between bands 1(3) and 2(4), which were plotted to ensure we understand which bandgap closings lead to changes in Chern numbers.\\

We find that the Chern numbers of the first and second bands in the lower frequency range are given as $(C_1, C_2) = (-1, 0)$. The gap between these two bands shrinks as $\omega_0$ increases and closes at $\omega_0 \approx 4.7$ eV, which is accompanied by a change of Chern numbers to $(C_1, C_2) = (0, -1)$. Both Chern numbers remain unchanged for the remainder of the investigated frequency interval. Chern number transitions also appear for bands 3 and 4. In this case, the Chern numbers start as $(C_3, C_4) = (-1, 1)$. Then the two bands approach each other until the bands cross near the $X$ point at driving frequency $\omega_0\approx 8.3$ eV. This gap closing is accompanied by a change in Chern number as $(C_3, C_4) = (-2, 2)$. A final band crossing occurs at driving frequency $\omega_0 \approx 9.3$ eV near the $M$ point and Chern numbers change to $(C_3, C_4) = (1, -1)$.

\begin{figure}[H]
    \centering
    \vspace{5pt}
    \includegraphics[width=\linewidth]{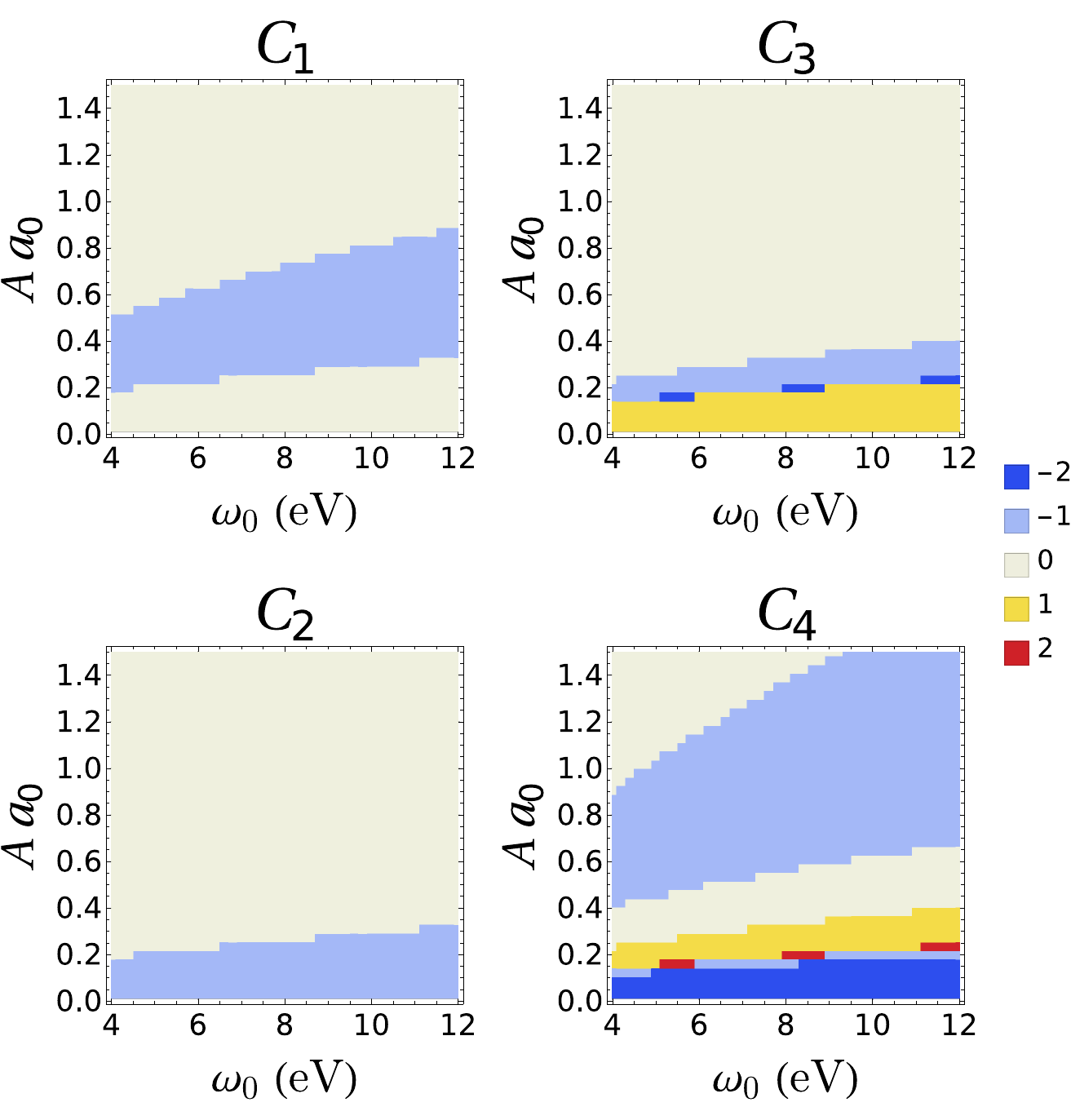}
    \caption{The phase diagram showing the transitions of the Chern number of band 1 to 4.}
    \label{fig:topo_phase_diag}
\end{figure}

To get a clearer picture of topology as a function of driving parameters, we also show a topological phase diagram in Fig. \ref{fig:topo_phase_diag}.

The phase diagram shows how the Chern number of each energy band changes against the various driving strengths and frequencies. From this figure, we can observe a plethora of topological transitions due to the influence of light. The electron bands even obtain a Chern number of $\pm2$ at certain values of $Aa_0$ and $\omega_0$. We also note that band 1 (4) has more valley Chern number transitions than bands 2 (3) since they can also interact with the band below (above).

\begin{widetext}
\begin{figure}[H] 
    \centering
    \includegraphics[width=0.48\textwidth]{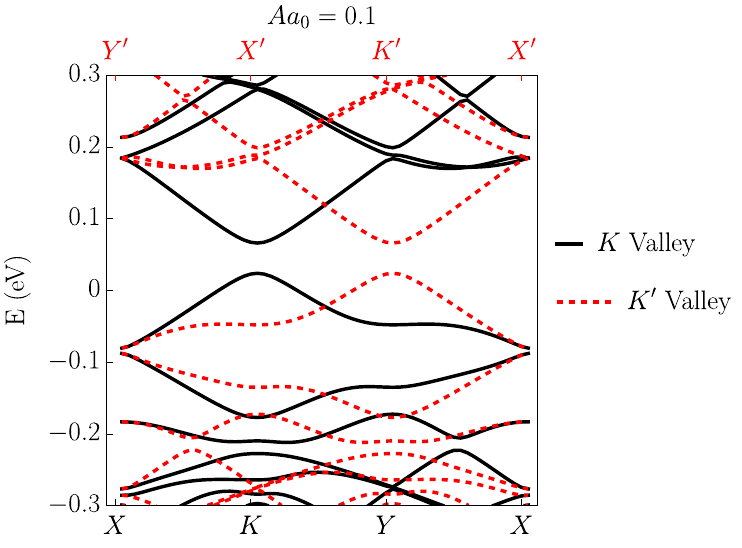}\includegraphics[width=0.48\textwidth]{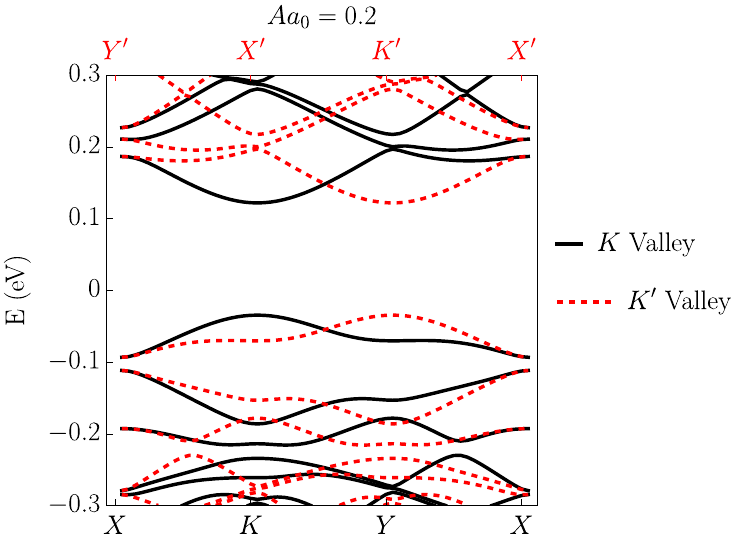}\\
    \includegraphics[width=0.48\textwidth]{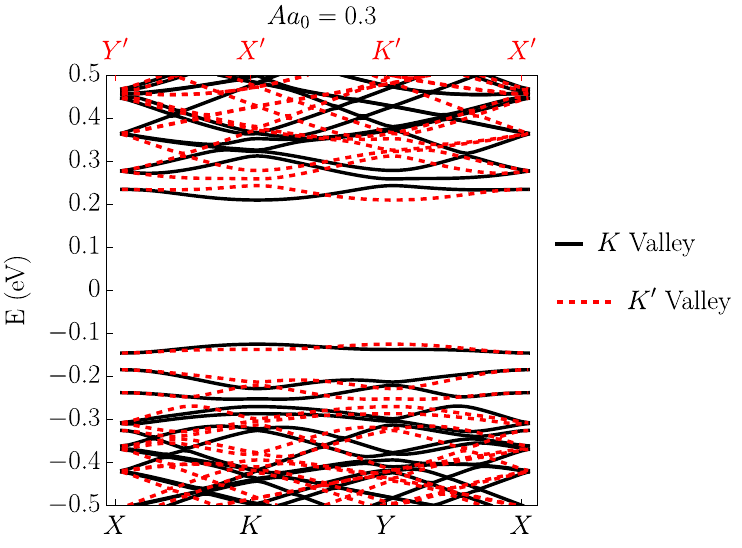}\includegraphics[width=0.48\textwidth]{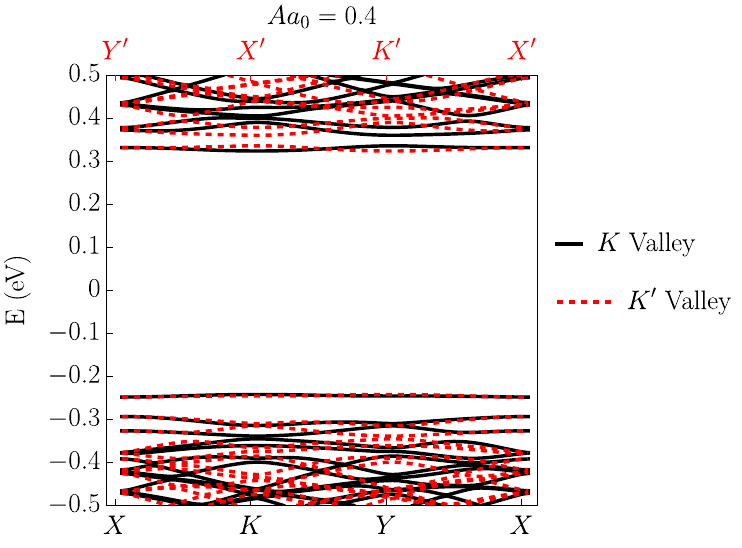}
    \caption{The quasi energy band structures of graphene/hBN system under the influence of circularly polarized light with different driving strength and frequency $\omega_0=2.1$ eV.}
    \label{Circularly polarized light band structure}
\end{figure}
\end{widetext}

Lastly, we investigate how changes in driving strengths $a_0A$ for a fixed driving frequency $\omega_0$ impact the quasi-energy bands in Fig. \ref{Circularly polarized light band structure}. 

Our result suggests that increasing the driving strength $A$ (with a fixed $\omega_0$) not only enlarges the gap between the hole and electron bands, but also flattens the bands. This feature is interesting since it indicates that circularly polarized light induces strong correlations between electrons in these bands.

\section{Conclusion}
\label{sec:concluscion}

In summary, we have studied a G-hBN hetero-bilayer under the influence of a circularly polarized light. We began by reviewing the equilibrium case and reproducing literature results. To account for the effect of circularly polarized light, we then introduced a vector potential using the minimal substitution method. Subsequently, we employed Floquet theory, which led us to our Floquet-Bloch Hamiltonian for our system Eq. (\ref{eq: Floquet Matrix Elements}, \ref{eq:Hg_Floquet}, \ref{eq:VhBN_Floquet}), which permitted a numerical treatment. We demonstrated that the application of circularly polarized light results in changes to the Chern numbers of four low-energy bands, providing a topological phase diagram. This result suggests that the application of circularly polarized light offers a promising approach to obtaining topological materials from the untwisted graphene-hBN system. Our observations may be beneficial for future electronic applications because they closely mirror results from more challenging-to-synthesize twisted Moiré materials.

For future work, several directions can be suggested. For instance, one could study this material under different types of light, such as twisted light or any arbitrary electromagnetic fields. One might also want to introduce the effect of magnetic fields, pressure, defects, and impurities, and study their interplay with periodic driving. It might also be helpful to study the effect of light on the twisted version of this material which is yet another type of moir\'{e} structure. Furthermore, the physical properties of this material could be further harnessed by developing simplified models in the interacting limit.
\section{Acknowledgements}
M.V. gratefully acknowledges the support provided by the Interdisciplinary Research Center(IRC) for Intelligent Secure Systems (ISS) at King Fahd University of Petroleum \& Minerals (KFUPM) for funding his contribution to this work through internal research grant No. INSS2507. 
\newpage
\bibliographystyle{unsrt}
\bibliography{ref}

\begin{thebibliography}{10}

\bibitem{Hentschel2001}
M.~Hentschel, R.~Kienberger, Ch. Spielmann, G.~A. Reider, N.~Milosevic,
  T.~Brabec, P.~Corkum, U.~Heinzmann, M.~Drescher, and F.~Krausz.
\newblock Attosecond metrology.
\newblock {\em Nature}, 414(6863):509--513, November 2001.

\bibitem{Hu2006}
S.~X. Hu and L.~A. Collins.
\newblock Attosecond pump probe: Exploring ultrafast electron motion inside an
  atom.
\newblock {\em Physical Review Letters}, 96(7):073004, February 2006.

\bibitem{Rana2009}
Dhanvir~Singh Rana, Iwao Kawayama, Krushna Mavani, Kouhei Takahashi, Hironaru
  Murakami, and Masayoshi Tonouchi.
\newblock Understanding the nature of ultrafast polarization dynamics of
  ferroelectric memory in the multiferroic bifeo3.
\newblock {\em Advanced Materials}, 21(28):2881--2885, July 2009.

\bibitem{Sheu2014}
Y.~M. Sheu, S.~A. Trugman, L.~Yan, Q.~X. Jia, A.~J. Taylor, and R.~P.
  Prasankumar.
\newblock Using ultrashort optical pulses to couple ferroelectric and
  ferromagnetic order in an oxide heterostructure.
\newblock {\em Nature Communications}, 5(1), December 2014.

\bibitem{Zhang2021}
Yuan Zhang, Junfeng Dai, Xiangli Zhong, Dongwen Zhang, Gaokuo Zhong, and
  Jiangyu Li.
\newblock Probing ultrafast dynamics of ferroelectrics by time‐resolved
  pump‐probe spectroscopy.
\newblock {\em Advanced Science}, 8(22), October 2021.

\bibitem{Chen2016}
F.~Chen, Y.~Zhu, S.~Liu, Y.~Qi, H.~Y. Hwang, N.~C. Brandt, J.~Lu, F.~Quirin,
  H.~Enquist, P.~Zalden, T.~Hu, J.~Goodfellow, M.-J. Sher, M.~C. Hoffmann,
  D.~Zhu, H.~Lemke, J.~Glownia, M.~Chollet, A.~R. Damodaran, J.~Park, Z.~Cai,
  I.~W. Jung, M.~J. Highland, D.~A. Walko, J.~W. Freeland, P.~G. Evans,
  A.~Vailionis, J.~Larsson, K.~A. Nelson, A.~M. Rappe, K.~Sokolowski-Tinten,
  L.~W. Martin, H.~Wen, and A.~M. Lindenberg.
\newblock Ultrafast terahertz-field-driven ionic response in ferroelectric
  batio3.
\newblock {\em Physical Review B}, 94(18):180104, November 2016.

\bibitem{Stich2013}
Dominik Stich, Florian Späth, Hannes Kraus, Andreas Sperlich, Vladimir
  Dyakonov, and Tobias Hertel.
\newblock Triplet–triplet exciton dynamics in single-walled carbon nanotubes.
\newblock {\em Nature Photonics}, 8(2):139--144, December 2013.

\bibitem{Bai2018}
Yusong Bai, Jean-Hubert Olivier, George Bullard, Chaoren Liu, and Michael~J.
  Therien.
\newblock Dynamics of charged excitons in electronically and morphologically
  homogeneous single-walled carbon nanotubes.
\newblock {\em Proceedings of the National Academy of Sciences},
  115(4):674--679, January 2018.

\bibitem{Birkmeier2022}
Konrad Birkmeier, Tobias Hertel, and Achim Hartschuh.
\newblock Probing the ultrafast dynamics of excitons in single semiconducting
  carbon nanotubes.
\newblock {\em Nature Communications}, 13(1), October 2022.

\bibitem{Jager2017}
Marieke~F. Jager, Christian Ott, Peter~M. Kraus, Christopher~J. Kaplan, Winston
  Pouse, Robert~E. Marvel, Richard~F. Haglund, Daniel~M. Neumark, and
  Stephen~R. Leone.
\newblock Tracking the insulator-to-metal phase transition in vo 2 with
  few-femtosecond extreme uv transient absorption spectroscopy.
\newblock {\em Proceedings of the National Academy of Sciences},
  114(36):9558--9563, August 2017.

\bibitem{Bionta2021}
Mina~R. Bionta, Elissa Haddad, Adrien Leblanc, Vincent Gruson, Philippe
  Lassonde, Heide Ibrahim, Jérémie Chaillou, Nicolas Émond, Martin~R. Otto,
  Álvaro Jiménez-Galán, Rui E.~F. Silva, Misha Ivanov, Bradley~J. Siwick,
  Mohamed Chaker, and François Légaré.
\newblock Tracking ultrafast solid-state dynamics using high harmonic
  spectroscopy.
\newblock {\em Physical Review Research}, 3(2):023250, June 2021.

\bibitem{Deb2022}
Marwan Deb, Elena Popova, Henri-Yves Jaffrès, Niels Keller, and Matias
  Bargheer.
\newblock Controlling high-frequency spin-wave dynamics using double-pulse
  laser excitation.
\newblock {\em Physical Review Applied}, 18(4):044001, October 2022.

\bibitem{Graves2013}
C.~E. Graves, A.~H. Reid, T.~Wang, B.~Wu, S.~de~Jong, K.~Vahaplar, I.~Radu,
  D.~P. Bernstein, M.~Messerschmidt, L.~Müller, R.~Coffee, M.~Bionta, S.~W.
  Epp, R.~Hartmann, N.~Kimmel, G.~Hauser, A.~Hartmann, P.~Holl, H.~Gorke, J.~H.
  Mentink, A.~Tsukamoto, A.~Fognini, J.~J. Turner, W.~F. Schlotter, D.~Rolles,
  H.~Soltau, L.~Strüder, Y.~Acremann, A.~V. Kimel, A.~Kirilyuk, Th. Rasing,
  J.~Stöhr, A.~O. Scherz, and H.~A. Dürr.
\newblock Nanoscale spin reversal by non-local angular momentum transfer
  following ultrafast laser excitation in ferrimagnetic gdfeco.
\newblock {\em Nature Materials}, 12(4):293--298, March 2013.

\bibitem{Stanciu2007}
C.~D. Stanciu, F.~Hansteen, A.~V. Kimel, A.~Kirilyuk, A.~Tsukamoto, A.~Itoh,
  and Th. Rasing.
\newblock All-optical magnetic recording with circularly polarized light.
\newblock {\em Physical Review Letters}, 99(4):047601, July 2007.

\bibitem{Rini2007}
Matteo Rini, Ra’anan Tobey, Nicky Dean, Jiro Itatani, Yasuhide Tomioka,
  Yoshinori Tokura, Robert~W. Schoenlein, and Andrea Cavalleri.
\newblock Control of the electronic phase of a manganite by mode-selective
  vibrational excitation.
\newblock {\em Nature}, 449(7158):72--74, September 2007.

\bibitem{Weidinger2017}
Simon~A. Weidinger and Michael Knap.
\newblock Floquet prethermalization and regimes of heating in a periodically
  driven, interacting quantum system.
\newblock {\em Scientific Reports}, 7(1), April 2017.

\bibitem{Giovannini2019}
Umberto~De Giovannini and Hannes Hübener.
\newblock Floquet analysis of excitations in materials.
\newblock {\em Journal of Physics: Materials}, 3(1):012001, October 2019.

\bibitem{Oka2019}
Takashi Oka and Sota Kitamura.
\newblock Floquet engineering of quantum materials.
\newblock {\em Annual Review of Condensed Matter Physics}, 10(1):387--408,
  March 2019.

\bibitem{Sentef2015}
M.A. Sentef, M.~Claassen, A.F. Kemper, B.~Moritz, T.~Oka, J.K. Freericks, and
  T.P. Devereaux.
\newblock Theory of floquet band formation and local pseudospin textures in
  pump-probe photoemission of graphene.
\newblock {\em Nature Communications}, 6(1), May 2015.

\bibitem{Ito2023}
S.~Ito, M.~Schüler, M.~Meierhofer, S.~Schlauderer, J.~Freudenstein,
  J.~Reimann, D.~Afanasiev, K.~A. Kokh, O.~E. Tereshchenko, J.~Güdde, M.~A.
  Sentef, U.~Höfer, and R.~Huber.
\newblock Build-up and dephasing of floquet–bloch bands on subcycle
  timescales.
\newblock {\em Nature}, 616(7958):696--701, April 2023.

\bibitem{Sentef2020}
Michael~A. Sentef, Jiajun Li, Fabian Künzel, and Martin Eckstein.
\newblock Quantum to classical crossover of floquet engineering in correlated
  quantum systems.
\newblock {\em Physical Review Research}, 2(3):033033, July 2020.

\bibitem{Kennes2019}
Dante~M. Kennes, Martin Claassen, Michael~A. Sentef, and Christoph Karrasch.
\newblock Light-induced d -wave superconductivity through floquet-engineered
  fermi surfaces in cuprates.
\newblock {\em Physical Review B}, 100(7):075115, August 2019.

\bibitem{Eckhardt2022}
Christian~J. Eckhardt, Giacomo Passetti, Moustafa Othman, Christoph Karrasch,
  Fabio Cavaliere, Michael~A. Sentef, and Dante~M. Kennes.
\newblock Quantum floquet engineering with an exactly solvable tight-binding
  chain in a cavity.
\newblock {\em Communications Physics}, 5(1), May 2022.

\bibitem{Topp2021}
Gabriel~E. Topp, Christian~J. Eckhardt, Dante~M. Kennes, Michael~A. Sentef, and
  Päivi Törmä.
\newblock Light-matter coupling and quantum geometry in moiré materials.
\newblock {\em Physical Review B}, 104(6):064306, August 2021.

\bibitem{VinasBostroem2020}
Emil Vinas~Boström, Martin Claassen, James McIver, Gregor Jotzu, Angel Rubio,
  and Michael Sentef.
\newblock Light-induced topological magnons in two-dimensional van der waals
  magnets.
\newblock {\em SciPost Physics}, 9(4), October 2020.

\bibitem{Kibis2020}
O.~V. Kibis, I.~V. Iorsh, and I.~A. Shelykh.
\newblock Floquet engineering of 2d materials.
\newblock {\em Journal of Physics: Conference Series}, 1461(1):012064, March
  2020.

\bibitem{Kong2022}
Xiangru Kong, Wei Luo, Linyang Li, Mina Yoon, Tom Berlijn, and Liangbo Liang.
\newblock Floquet band engineering and topological phase transitions in 1t’
  transition metal dichalcogenides.
\newblock {\em 2D Materials}, 9(2):025005, January 2022.

\bibitem{Rudner2020}
Mark~S. Rudner and Netanel~H. Lindner.
\newblock Band structure engineering and non-equilibrium dynamics in floquet
  topological insulators.
\newblock {\em Nature Reviews Physics}, 2(5):229--244, May 2020.

\bibitem{Castro2022}
Alberto Castro, Umberto De~Giovannini, Shunsuke~A. Sato, Hannes Hübener, and
  Angel Rubio.
\newblock Floquet engineering the band structure of materials with optimal
  control theory.
\newblock {\em Physical Review Research}, 4(3):033213, September 2022.

\bibitem{Fleury2016}
Romain Fleury, Alexander~B Khanikaev, and Andrea Alù.
\newblock Floquet topological insulators for sound.
\newblock {\em Nature Communications}, 7(1), June 2016.

\bibitem{Zhan2023}
Fangyang Zhan, Junjie Zeng, Zhuo Chen, Xin Jin, Jing Fan, Tingyong Chen, and
  Rui Wang.
\newblock Floquet engineering of nonequilibrium valley-polarized quantum
  anomalous hall effect with tunable chern number.
\newblock {\em Nano Letters}, 23(6):2166--2172, March 2023.

\bibitem{Shin2018}
Dongbin Shin, Hannes Hübener, Umberto De~Giovannini, Hosub Jin, Angel Rubio,
  and Noejung Park.
\newblock Phonon-driven spin-floquet magneto-valleytronics in mos2.
\newblock {\em Nature Communications}, 9(1), February 2018.

\bibitem{Cao2024}
Haijun Cao, Jia-Tao Sun, and Sheng Meng.
\newblock Floquet engineering of anomalous hall effects in monolayer mos2.
\newblock {\em npj Quantum Materials}, 9(1), November 2024.

\bibitem{Grushin2014}
Adolfo~G. Grushin, Álvaro Gómez-León, and Titus Neupert.
\newblock Floquet fractional chern insulators.
\newblock {\em Physical Review Letters}, 112(15):156801, April 2014.

\bibitem{Ashida2020}
Yuto Ashida, Ataç İmamoğlu, Jérôme Faist, Dieter Jaksch, Andrea Cavalleri,
  and Eugene Demler.
\newblock Quantum electrodynamic control of matter: Cavity-enhanced
  ferroelectric phase transition.
\newblock {\em Physical Review X}, 10(4):041027, November 2020.

\bibitem{Li2020}
Yantao Li, H.~A. Fertig, and Babak Seradjeh.
\newblock Floquet-engineered topological flat bands in irradiated twisted
  bilayer graphene.
\newblock {\em Physical Review Research}, 2(4):043275, November 2020.

\bibitem{Topp2019}
Gabriel~E. Topp, Gregor Jotzu, James~W. McIver, Lede Xian, Angel Rubio, and
  Michael~A. Sentef.
\newblock Topological floquet engineering of twisted bilayer graphene.
\newblock {\em Physical Review Research}, 1(2):023031, September 2019.

\bibitem{Vogl_2020}
Michael Vogl, Martin Rodriguez-Vega, and Gregory~A. Fiete.
\newblock Effective floquet hamiltonians for periodically driven twisted
  bilayer graphene.
\newblock {\em Physical Review B}, 101(23), June 2020.

\bibitem{Vogl2021}
Michael Vogl, Martin Rodriguez-Vega, Benedetta Flebus, Allan~H. MacDonald, and
  Gregory~A. Fiete.
\newblock Floquet engineering of topological transitions in a twisted
  transition metal dichalcogenide homobilayer.
\newblock {\em Physical Review B}, 103(1):014310, January 2021.

\bibitem{Mitrano2016}
M.~Mitrano, A.~Cantaluppi, D.~Nicoletti, S.~Kaiser, A.~Perucchi, S.~Lupi,
  P.~Di~Pietro, D.~Pontiroli, M.~Riccò, S.~R. Clark, D.~Jaksch, and
  A.~Cavalleri.
\newblock Possible light-induced superconductivity in k3c60 at high
  temperature.
\newblock {\em Nature}, 530(7591):461--464, February 2016.

\bibitem{Fausti2011}
D.~Fausti, R.~I. Tobey, N.~Dean, S.~Kaiser, A.~Dienst, M.~C. Hoffmann, S.~Pyon,
  T.~Takayama, H.~Takagi, and A.~Cavalleri.
\newblock Light-induced superconductivity in a stripe-ordered cuprate.
\newblock {\em Science}, 331(6014):189--191, January 2011.

\bibitem{Suda2015}
Masayuki Suda, Reizo Kato, and Hiroshi~M. Yamamoto.
\newblock Light-induced superconductivity using a photoactive electric double
  layer.
\newblock {\em Science}, 347(6223):743--746, February 2015.

\bibitem{Fava2024}
S.~Fava, G.~De~Vecchi, G.~Jotzu, M.~Buzzi, T.~Gebert, Y.~Liu, B.~Keimer, and
  A.~Cavalleri.
\newblock Magnetic field expulsion in optically driven yba2cu3o6.48.
\newblock {\em Nature}, 632(8023):75--80, July 2024.

\bibitem{Nova2019}
T.~F. Nova, A.~S. Disa, M.~Fechner, and A.~Cavalleri.
\newblock Metastable ferroelectricity in optically strained srtio 3.
\newblock {\em Science}, 364(6445):1075--1079, June 2019.

\bibitem{Li2019}
Xian Li, Tian Qiu, Jiahao Zhang, Edoardo Baldini, Jian Lu, Andrew~M. Rappe, and
  Keith~A. Nelson.
\newblock Terahertz field–induced ferroelectricity in quantum paraelectric
  srtio 3.
\newblock {\em Science}, 364(6445):1079--1082, June 2019.

\bibitem{Nova2016}
T.~F. Nova, A.~Cartella, A.~Cantaluppi, M.~Först, D.~Bossini, R.~V.
  Mikhaylovskiy, A.~V. Kimel, R.~Merlin, and A.~Cavalleri.
\newblock An effective magnetic field from optically driven phonons.
\newblock {\em Nature Physics}, 13(2):132--136, October 2016.

\bibitem{McIver2019}
J.~W. McIver, B.~Schulte, F.-U. Stein, T.~Matsuyama, G.~Jotzu, G.~Meier, and
  A.~Cavalleri.
\newblock Light-induced anomalous hall effect in graphene.
\newblock {\em Nature Physics}, 16(1):38--41, November 2019.

\bibitem{5k9m-mfbz}
Miftah Hadi~Syahputra Anfa, Sabri Elatresh, Hocine Bahlouli, and Michael Vogl.
\newblock Effective $k$-valley hamiltonian for transition metal dichalcogenide
  bilayers under pressure and application to twisted bilayers with
  pressure-induced topological phase transitions.
\newblock {\em Phys. Rev. B}, 111:245434, Jun 2025.

\bibitem{Moon_2014}
Pilkyung Moon and Mikito Koshino.
\newblock Electronic properties of graphene/hexagonal-boron-nitride moiré
  superlattice.
\newblock {\em Physical Review B}, 90(15), October 2014.

\bibitem{Koshino2013}
Mikito Koshino and Tsuneya Ando.
\newblock {\em Electronic Properties of Monolayer and Multilayer Graphene},
  pages 173--211.
\newblock Springer International Publishing, December 2013.

\bibitem{Liu2003}
Lei Liu, Y.~P. Feng, and Z.~X. Shen.
\newblock Structural and electronic properties of h-bn.
\newblock {\em Physical Review B}, 68(10):104102, September 2003.

\bibitem{Slawinska2010}
J.~Sławińska, I.~Zasada, and Z.~Klusek.
\newblock Energy gap tuning in graphene on hexagonal boron nitride bilayer
  system.
\newblock {\em Physical Review B}, 81(15):155433, April 2010.

\bibitem{bloch1929}
Felix Bloch.
\newblock {\"U}ber die quantenmechanik der elektronen in kristallgittern.
\newblock {\em Zeitschrift f{\"u}r physik}, 52(7-8):555--600, 1929.

\bibitem{rudner2013}
Mark~S Rudner, Netanel~H Lindner, Erez Berg, and Michael Levin.
\newblock Anomalous edge states and the bulk-edge correspondence for
  periodically driven two-dimensional systems.
\newblock {\em Physical Review X}, 3(3):031005, 2013.

\bibitem{Wawer_2021}
Lukas Wawer and Michael Fleischhauer.
\newblock Chern number and berry curvature for gaussian mixed states of
  fermions.
\newblock {\em Physical Review B}, 104(9), September 2021.

\bibitem{Fukui_2005}
Takahiro Fukui, Yasuhiro Hatsugai, and Hiroshi Suzuki.
\newblock Chern numbers in discretized brillouin zone: Efficient method of
  computing (spin) hall conductances.
\newblock {\em Journal of the Physical Society of Japan}, 74(6):1674–1677,
  June 2005.

\end{thebibliography}
\end{document}